\def\papertitle{Perceptually Motivated Alignment and Interpolation of Pitch-Aligned Time-Frequency Representations}
\def\paperauthorA{Shahan Nercessian}
\def\paperauthorB{Jeff Sontag}
\def\paperauthorC{Alejandro Koretzky}
\documentclass[twoside,a4paper]{article}
\usepackage{etoolbox}

\usepackage[print]{dafx26v3}

\usepackage{amsmath,amssymb,amsfonts,amsthm}
\usepackage{siunitx}
\usepackage{euscript}
\usepackage[T1]{fontenc}
\usepackage[utf8]{inputenc}
\usepackage{ifpdf}
\usepackage[english]{babel}
\usepackage{caption}
\usepackage{subfig} % or can use subcaption package
\usepackage{color}
\usepackage{booktabs}
\usepackage{lipsum}
\usepackage{bm}

\input glyphtounicode
\ninept

\newcounter{numauth}
\newcounter{listcnt}
\newcommand\authcnt[1]{\ifdefined#1 \stepcounter{numauth} \fi}

\newcommand\addauth[1]{
\ifdefined#1 
\stepcounter{listcnt}
\ifnum \value{listcnt}<\value{numauth}
\appto\authorslist{, #1}
\else
\appto\authorslist{~and~#1}
\fi
\fi}
\authcnt{\paperauthorB}
\authcnt{\paperauthorC}
\authcnt{\paperauthorD}
\authcnt{\paperauthorE}
\authcnt{\paperauthorF}
\authcnt{\paperauthorG}
\authcnt{\paperauthorH}
\authcnt{\paperauthorI}
\authcnt{\paperauthorJ}
\def\authorslist{\paperauthorA}
\addauth{\paperauthorB}
\addauth{\paperauthorC}
\addauth{\paperauthorD}
\addauth{\paperauthorE}
\addauth{\paperauthorF}
\addauth{\paperauthorG}
\addauth{\paperauthorH}
\addauth{\paperauthorI}
\addauth{\paperauthorJ}
\DeclareMathOperator{\diag}{diag}
\usepackage{algorithm}
\usepackage[noend]{algpseudocode}

\usepackage{times}
\newif\ifpdf
\ifx\pdfoutput\relax
\else
   \ifcase\pdfoutput
      \pdffalse
   \else
      \pdftrue
   \fi
\fi

\ifpdf % compiling with pdflatex
  \usepackage[pdftex,
    pdftitle={\papertitle},
    pdfauthor={\authorslist},
    pdfsubject={Proceedings of the 29th International Conference on Digital Audio Effects (DAFx26)},
    colorlinks=false, % links are activated as color boxes instead of color text
    bookmarksnumbered, % use section numbers with bookmarks
    pdfstartview=XYZ % start with zoom=100% instead of full screen; especially useful if working with a big screen :-)
  ]{hyperref}
  \usepackage[pdftex]{graphicx}
\else % compiling with latex
  \usepackage[dvips]{epsfig,graphicx}
  \usepackage[dvips,
    pdftitle={\papertitle},
    pdfauthor={\authorslist},
    pdfsubject={Proceedings of the 29th International Conference on Digital Audio Effects (DAFx26)},
    colorlinks=false, % no color links
    bookmarksnumbered, % use section numbers with bookmarks
    pdfstartview=XYZ % start with zoom=100% instead of full screen
  ]{hyperref}
\fi
\usepackage[hypcap=true]{caption}
\title{\papertitle}

\affiliation
{\paperauthorA, \paperauthorB, and \paperauthorC}
{\href{https://www.splice.com}{Splice} \\ New York, USA\\
{\tt \{{\href{mailto:shahan.nercessian@splice.com}{shahan.nercessian}}|{\href{mailto:jeff.sontag@splice.com}{jeff.sontag}|{\href{mailto:alejandro.koretzky@splice.com}{alejandro.koretzky}}\}@splice.com}}
}

\begin{document}
% more pdf-tex settings:
\ifpdf % used graphic file format for pdflatex
  \DeclareGraphicsExtensions{.png,.jpg,.pdf}
\else  % used graphic file format for latex
  \DeclareGraphicsExtensions{.eps}
\fi

%\makeatletter
%\pdfbookmark[0]{\@pdftitle}{title}
%\makeatother

\maketitle

\begin{abstract}
We propose a framework for the alignment and interpolation of pitch-aligned time-frequency representations. Building on the tonal interval vector, we introduce a series of extensions that reformulate it as an invertible operator, culminating in a new feature extractor that embeds perceptual consonance priors within a pitch-aligned representation. Accordingly, we develop methods for aligning and interpolating between musical structures. For alignment, we cast the problem as a permutation search under perceptually weighted distances, enabling robust matching without explicit key detection, which is inherently ambiguous in music information retrieval. For interpolation, we employ optimal transport to generate musically meaningful transitions between pitch distributions, proposing a circular formulation over the Circle of Fourths/Fifths to respect harmonic structure. Finally, we learn a low-dimensional geometric representation of scale structure aimed to factorize scale color and density. Experimental results demonstrate the effectiveness and musicality of the proposed methods. Together, our contributions provide a principled, purely signal processing approach to modeling pitch structure in audio and symbolic music signals.
\end{abstract}

\newcommand{\norm}[1]{\left\lVert#1\right\rVert}

\section{Introduction}
\label{sec:intro}
A longstanding objective in music information retrieval and audio signal processing is the development of pitch-aligned time-frequency representations (PTFRs) describing pitch content in audio and symbolic music signals. Representations ranging from constant-Q transforms and their derived chroma and pitch class profile (PCP) features \cite{gomez2006tonal}, to transcription-based piano rolls and pitch activation maps \cite{BasicPitch}, provide structured views of pitched material at varying levels of abstraction. PTFRs underpin a wide range of tasks, including chord recognition and key estimation \cite{muller2015fundamentals}.

Beyond analysis, these representations invite operations such as alignment, interpolation, and general transformation of pitched material, enabling re-harmonization and cross-context comparison. However, existing methods are limited in scope, or may rely on standard Euclidean formulations \cite{ellis2007identifying} that do not fully reflect perceptual consonance or harmonic proximity, yielding outputs that are mathematically consistent but musically implausible. The perceptually motivated tonal interval vector (TIV) \cite{TIVorig} offer a compelling alternative, but has been used primarily for analysis purposes \cite{TIVdafx}. These limitations become apparent when manipulating pitch structure across musical contexts. As musical phrases derive their identity through their relationship to harmonic context, instrumentation, and surrounding material, it begs the problem of how to preserve their structure while maintaining perceptual coherence. Adapting a chord progression to a new melody, modulating between tonalities, or aligning independently generated layers all require development of a ``toolkit'' that is musically relevant. These tasks require mathematical tools that enable such adaptation while preserving perceptually meaningful musical relationships.

Recent advances in neural audio synthesis have further amplified the importance of structured manipulation of pitch content. Systems such as MusicGen \cite{copet2023simple} and diffusion-based approaches \cite{DiffARiff} demonstrate high-quality musical generation, but typically offer limited tools to control harmonic structure. In practice, this makes it difficult to perform operations such as re-harmonization, cross-key adaptation, and coherent alignment between independently generated musical layers. Building on these advances, there is growing interest to bridge the gap between high-level musical descriptions and low-level audio manipulations. This, in turn, motivates the development of methods that enable more fine-grained and perceptually meaningful manipulation of pitch content, both as standalone tools for analysis and composition, as well as interfaces for enriching generative models. Such methods can unlock the potential of generative systems to perform structured transformations of pitch content beyond the realm of standard signal processing.

In this work, we introduce perceptually motivated methods for aligning and interpolating PTFRs. Expanding on the TIV formulation, we develop a series of extensions reformulating it as an invertible operator, enabling its application to arbitrary-dimensional pitch representations while preserving its correlation to perceptual consonance. This gives rise to a means for alignment by determining optimal permutations under perceptually weighted distances. For interpolation, we build on optimal transport (OT) techniques \cite{OTdafx}, proposing a circular extension for chroma vectors along the Circle of Fourths/Fifths (CoF) \cite{levy1985theory}. Finally, we introduce a geometric scale representation providing an interpretable low-dimensional organization of pitch distributions that factorizes into root, density, and color components. These contributions establish a principled signal processing approach to modeling and manipulating pitch structure in both audio and symbolic music, with implications for conditional generative music systems.

This paper is organized as follows: Section~\ref{sec:tiv} overviews the TIV; Section~\ref{sec:extensions} introduces our proposed TIV extensions; Section~\ref{sec:alignment} applies these extensions to PTFR alignment; Section~\ref{sec:interpolation} explores musical interpolation; Section~\ref{sec:scale} develops a geometric scale representation; Section~\ref{sec:results} reports experimental results; Section~\ref{sec:conclusions} draws conclusions and outlines future work.

\begin{figure*}[hbt]
\centering
  \begin{minipage}{.3\textwidth}
	\centering
  \centerline{\includegraphics[width=0.95\columnwidth]{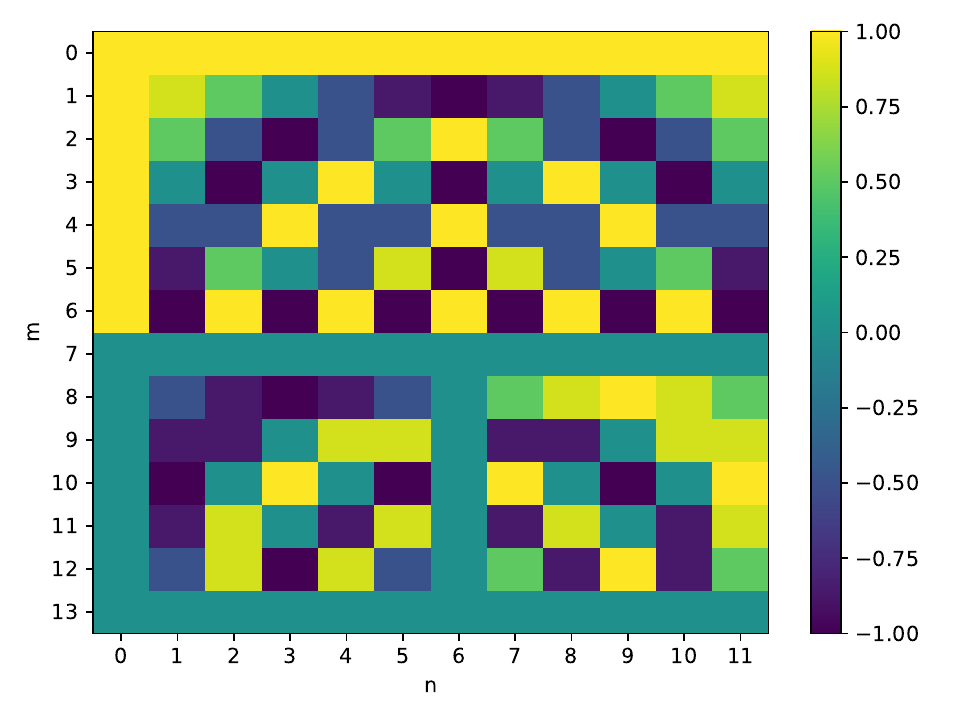}}
	\vspace{0.0em}(a)
  \end{minipage}
  \begin{minipage}{.3\textwidth}
	\centering
  \centerline{\includegraphics[width=0.95\columnwidth]{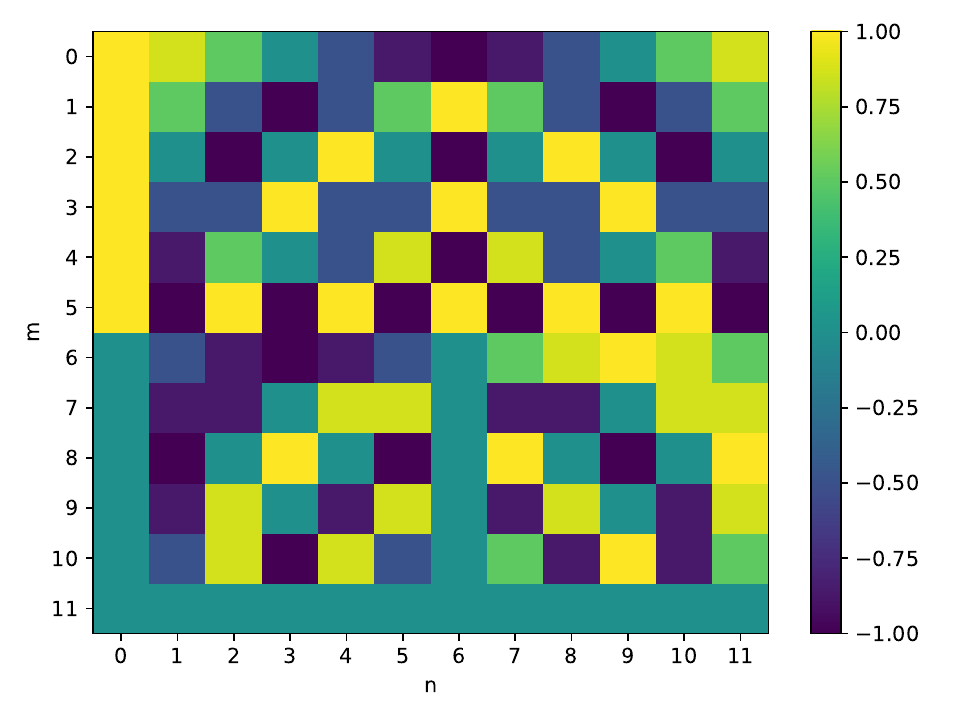}}
	\vspace{0.0em}(b)
  \end{minipage}
  \begin{minipage}{.3\textwidth}
	\centering
  \centerline{\includegraphics[width=0.95\columnwidth]{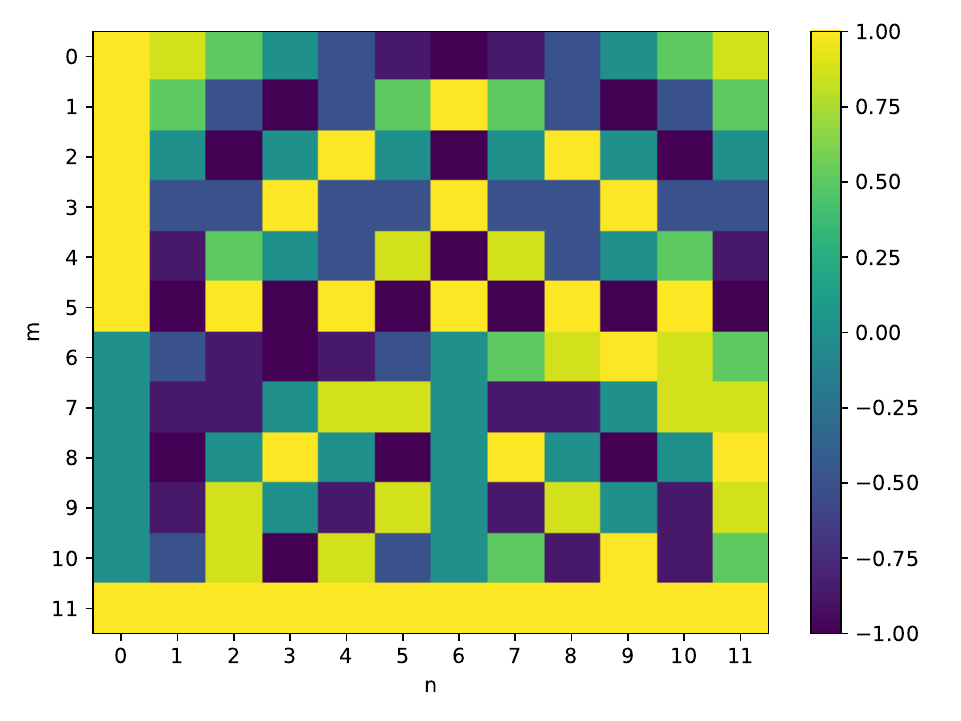}}
	\vspace{0.0em}(c)
  \end{minipage}
\caption{(a) $\bm{F_\text{real}}$, (b) $\bm{F}$ implicitly used by equation~(\ref{eqn:tiv}) in \cite{TIVorig, TIVdafx}, (c) proposed invertible $\bm{F}$.}
\vspace{-1.0em}
\label{fig:fourier}
\end{figure*}

\section{Tonal Interval Vectors}
\label{sec:tiv}

Given a chroma vector $\bm{c} \in \mathbb{R}^{N}$ (with $N=12$), its corresponding TIV $T(\bm{c}) \in \mathbb{C}^{6}$ is defined in \cite{TIVorig, TIVdafx} as

\begin{equation}
T(\bm{c})[k]= \bm{w_{*}}[k] \sum_{n=0}^{N-1} \bar{\bm{c}}[n] \exp{\frac{-j2\pi  kn}{N}}, \quad ~1\le k\le 6
\label{eqn:tiv}
\end{equation}
where $w_* \in \mathbb{R}^6$ is a vector of perceptual weights, and

\begin{equation}
\bar{\bm{c}}[n] = \frac{\bm{c}[n]}{\sum_{n=0}^{N-1} \bm{c}[n]}
\label{eqn:norm}
\end{equation}
is used to ensure an amplitude-invariant tonal representation. Equation~(\ref{eqn:tiv}) constitutes a Discrete Fourier Transform (DFT) analyzing the presence of different ``frequencies'' (i.e., tonal intervals) in the real-valued $\bar{c}$, discarding its DC component (equal to 1 due to the normalization in equation~(\ref{eqn:norm})), and weighting the result by $w_*$.

The motivation for TIVs is the assertion that for two arbitrary chroma vectors $\bm{x}$ and $\bm{y}$, a distance metric defined in the TIV space (i.e., $\norm{T(\bm{x}) - T(\bm{y})}_2$) corresponds better to the perception of consonance than the same metric defined in the chroma space (i.e., $\norm{\bm{x} - \bm{y}}_2$), due to the combination of normalization and weighting that it applies. The weights were derived from empirical ratings of dyad consonance used to adjust the contribution of each dimension in the space. An initial study determined a weight set for symbolic input as $\bm{w}_\text{symbolic}=[2, 11, 17, 16, 19, 7]$, using a brute-force search over a set of bounded integers \cite{TIVorig}. A follow-up \cite{TIVdafx} suggested a set of weights for audio input as $\bm{w}_\text{audio} = [3, 8, 11.5, 15, 14.5, 7.5]$. TIV extraction can be extended to a time-varying chromagram $\bm{C} \in \mathbb{R}^{12 \times M}$, applying equation~(\ref{eqn:tiv}) to the chroma vector at each time frame $m$, yielding $T(\bm{C}) \in \mathbb{C}^{6 \times M}$. Alternatively, we can time-average $\bm{C}$, yielding the PCP $\bm{c}_\text{PCP}$ \cite{gomez2006tonal} with TIV $T(\bm{c}_\text{PCP})$.
\begin{figure*}[hbt]
\centering
  \begin{minipage}{.40\textwidth}
	\centering
  \centerline{\includegraphics[width=0.95\columnwidth]{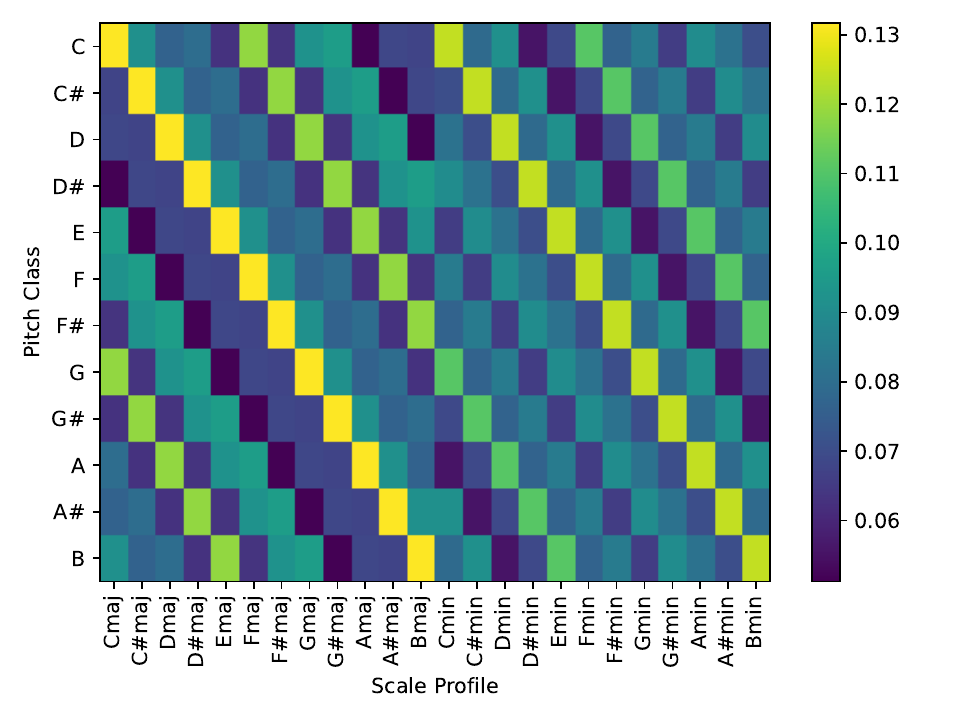}}
	\vspace{0.0em}(a)
  \end{minipage}
  \begin{minipage}{.40\textwidth}
	\centering
  \centerline{\includegraphics[width=0.95\columnwidth]{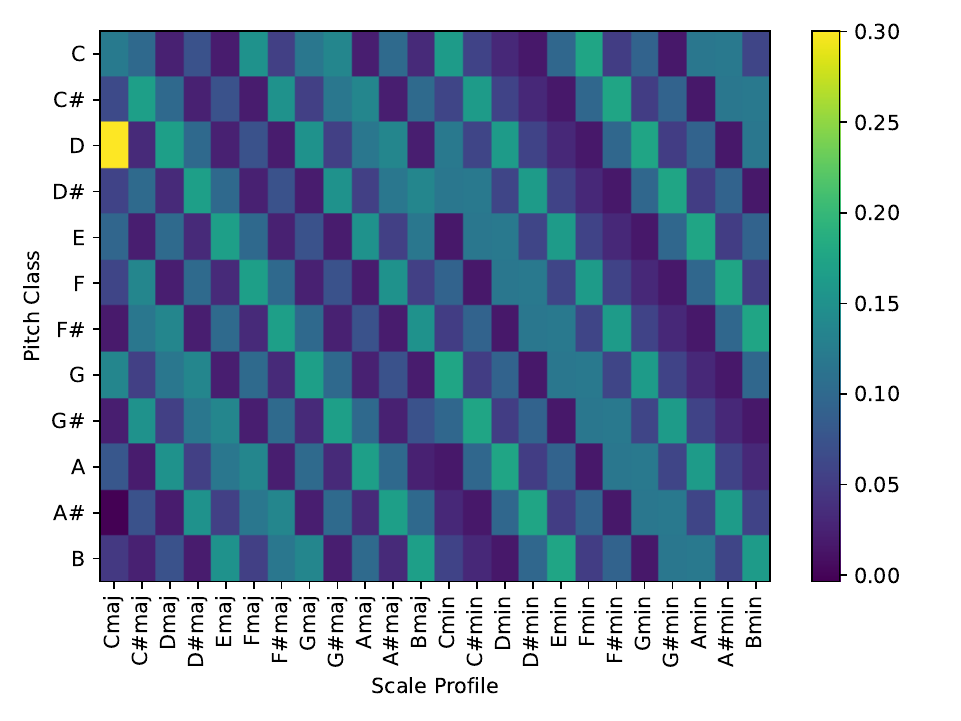}}
	\vspace{0.0em}(b)
  \end{minipage}
\caption{(a) Shaath and (b) Temperley pitch class profiles recovered through our proposed inverse filtering of TIVs provided in \texttt{TIV.lib} \cite{TIVdafx}, indicating a potentially problematic C major Temperley profile.}
\vspace{-1.0em}
\label{fig:profiles}
\end{figure*}

\section{TIV Extensions}
\label{sec:extensions}
We propose extensions to TIVs in order to support various applications in this work. We express the DFT in equation~(\ref{eqn:tiv}) as a matrix  $\bm{F}_\text{complex} \in \mathbb{C}^{(N / 2 + 1) \times N}$ for even $N$ and indices $0\le k\le N / 2$. This matrix has a real analog $\bm{F}_\text{real} = [\text{Re}({\bm{F}_\text{complex}}) ; \text{Im}({\bm{F}_\text{complex}})] \in \mathbb{R}^{(N + 2) \times N}$. Thus, $\bm{F}_\text{complex}\bm{c} \in \mathbb{C}^{N / 2 + 1}$ and $\bm{F}_\text{real}\bm{c} \in \mathbb{R}^{N + 2}$ contain precisely the same information, with the latter stacking the real and imaginary components of the former. We refer to stacking/unstacking of this form as \texttt{im2re}/\texttt{re2im}, respectively, with $\bm{F}_\text{real}\bm{c} = \texttt{im2re}(\bm{F}_\text{complex}\bm{c})$) and $\bm{F}_\text{complex}\bm{c} = \texttt{re2im}(\bm{F}_\text{real}\bm{c})$.

\begin{figure*}[hbt]
\centering
  \begin{minipage}{.40\textwidth}
	\centering
  \centerline{\includegraphics[width=0.95\columnwidth]{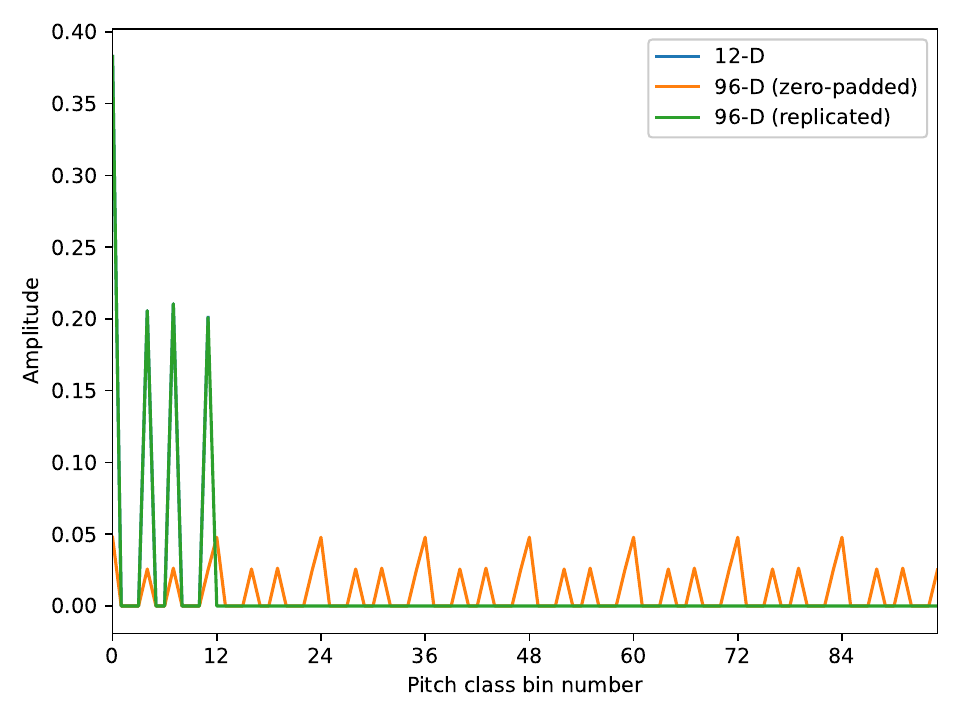}}
	\vspace{0.0em}(a)
  \end{minipage}
  \begin{minipage}{.40\textwidth}
	\centering
  \centerline{\includegraphics[width=0.95\columnwidth]{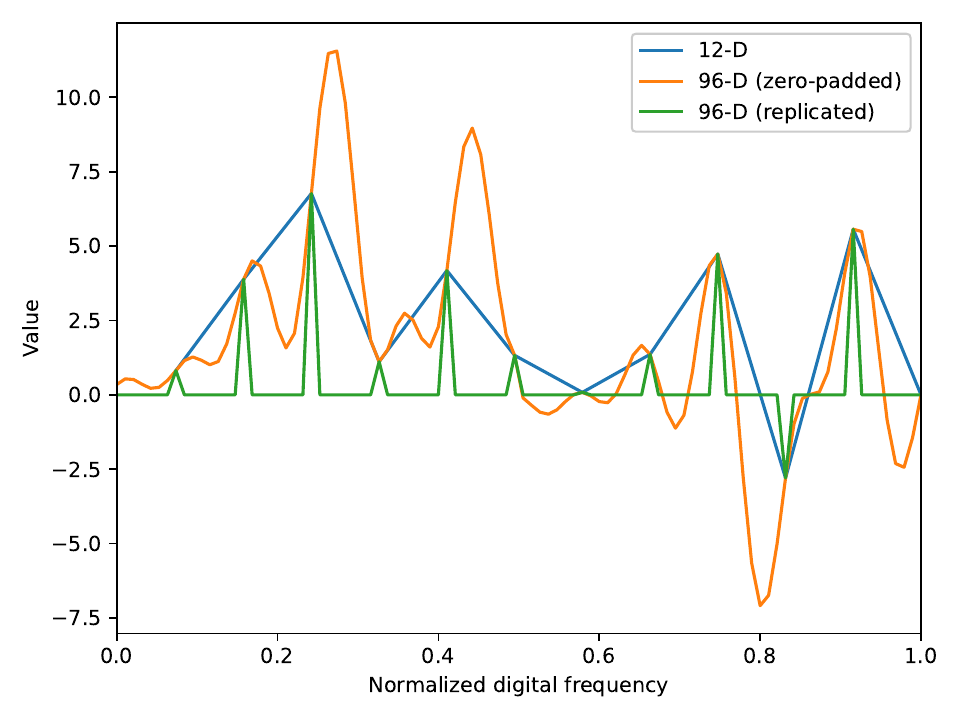}}
	\vspace{0.0em}(b)
  \end{minipage}
\caption{(a) $12$- and $96$-D PTFRs and (b) their corresponding TIVs using our proposed extensions.}
\vspace{-1.5em}
\label{fig:extension}
\end{figure*}

\subsection{Inversion}\label{ssec:inversion}

Despite having $N+2$ rows, $\bm{F}_\text{real}$ has rank $N$, as the DC and Nyquist frequency components correspond to purely real basis vectors whose imaginary parts are zero. Discarding the DC component in equation~(\ref{eqn:tiv}) breaks invertibility by removing a rank-contributing row of $\bm{F}_\text{real}$, reducing the rank of the resulting matrix from $N$ to $N-1$. We reformulate the transform in a manner that remains equivalent to equation~(\ref{eqn:tiv}) while preserving invertibility. We consider a basis $\bm{F} \in \mathbb{R}^{N \times N}$ that starts with $\bm{F}_\text{real}$, removes its two null components which do not contribute to its matrix rank to create a square matrix, and moves the DC component to the final row of the matrix. By construction, the resulting $\bm{F} \in \mathbb{R}^{N \times N}$ has full rank and is therefore invertible. Figure~\ref{fig:fourier} contrasts the different underlying Fourier matrices considered here. Next, we define the augmented vector $\bm{w}_{*,2\times}=[\bm{w}_* ; \bm{w}_{*,:-1}; \kappa]$, where $\bm{w}_{*,:-1} = [w_0, ..., w_{N / 2 - 2}]$ and $\kappa$ is a free parameter. We construct $\bm{W} = \diag{(\bm{w_{*,2\times}})}$ and the transform $\bm{T}=\bm{WF}$. With $\kappa=0$, equation~(\ref{eqn:tiv}) is recovered exactly via $T(\bm{c})=\texttt{re2im}(\bm{T}\bar{\bm{c}})$. This equivalence arises from a reparameterization in which the zero-valued component in equation~(\ref{eqn:tiv})—previously enforced by the structure of the Fourier basis—is now enforced via $\kappa=0$.

For $\kappa=0$, the TIV transform itself is not invertible, but its basis $\bm{F}$ is. This fact alone motivates a new PTFR that retains the perceptual smoothing effect of TIV while remaining aligned with the musical pitch axis (see Section~\ref{ssec:perceptual}). With $\kappa = 1$, we effectively retain the DC component of the input, whose value is equal to 1 by construction for $\bm{\bar{c}}$ but is not necessarily so if we also allow ourselves to apply the transform to $\bm{c}$ directly. In this setting, the generalized TIV transform becomes invertible, enabling inverse filtering via spectral division to recover the underlying chroma vector. We observe that doing so reveals inconsistencies in the \texttt{TIV.lib} implementation \cite{TIVdafx} of the Temperley profiles \cite{temperley1999whats}, as illustrated in Figure~\ref{fig:profiles}. While the Shaath profiles encode major and minor scale templates and their transpositions, the recovered Temperley profile exhibits an unexpected emphasis on D in the case of C major. This behavior may explain the degraded performance of these profiles as templates for key detection in our internal evaluations.

\subsection{Normalization}\label{ssec:norm}
We propose optional normalization steps applied to $\bm{F}$ and $w*$ (i.e., $\bm{W}$). Although $\bm{F}$ defines an invertible, purely real Fourier basis with orthogonal rows, these rows are not orthonormal as constructed, except for vectors with no imaginary components in their corresponding complex representation. We can normalize the rows of $\bm{F}$ to make it orthogonal, ensuring $\norm{\bm{Fx}-\bm{Fy}}_2 = \norm{\bm{x}-\bm{y}}_2$, which will be useful in Section~\ref{sec:alignment}. We correspondingly rescale the elements of $\bm{W}$ to maintain the previously established relationship between TIV distances and empirical consonance ratings in \cite{TIVorig}.

Moreover, the coefficients in $\bm{w_*}$ span the range of $[2, 19]$, which can inflate the magnitude of the TIVs by an order of magnitude (and therefore distance metrics defined in terms of them) relative to their respective chroma vectors. We propose an optional normalization to the weights $\bm{w}_{*,1:} = [w_1, ..., w_{N / 2 - 1}]$ such that they sum to $N / 2 - 1$, i.e., the sum of the corresponding weights had they been set to unity. This maintains $\kappa$ as a free parameter while making $\norm{\bm{Tx}-\bm{Ty}}_2 \approx \norm{\bm{x}-\bm{y}}_2$ in the relative sense. Together with the previous normalization steps, this facilitates more interpretable visualization and comparison of the perceptually smoothed PTFR introduced in Section~\ref{ssec:perceptual}.

\subsection{Higher-dimensional PTFRs}\label{ssec:PTFRs}

Next, we extend the TIV space defined strictly for $N=12$ to any $N=12o$, where $o$ is a non-negative integer indicating the number of octaves captured by a given PTFR. Such an extension analyzes tonal interval patterns extending beyond a single octave, foregoing octave equivalence captured by a chroma vector over when $o > 1$.

Our construction of invertible, optionally orthogonal $\bm{F}$, and our optional normalizations scale naturally to $12o$-D PTFRs since they are generally expressed in terms of $N$. Next, we must extend a baseline $\bm{w}_*$ for arbitrary $o > 1$. Since  a $12o$-point transform increases its tonal interval resolution by a factor $o$ relative to its $12$-point counterpart, we opt to oversample a canonical $\bm{w}_*$ using linear interpolation. The boundedness and monotonicity properties of linear interpolation ensure that the resulting coefficients remain non-negative. The tonal interval basis frequencies for $N=12$ will also be present precisely for any $N=12o$, with their corresponding weights preserved under our implementation. When we consider non-zero values for $\kappa$, a design choice arises. One natural approach is to interpolate the weights between $\kappa$ and their canonical $\bm{w}_{*,1}$ across the oversampling interval. Alternatively, we can interpolate them between 0 and their canonical $\bm{w}_{*,1}$, followed by the explicit assignment $\bm{w_{*,0}} \leftarrow \kappa$. We adopt the latter to remain consistent with the original DC-invariant TIV formulation, while enabling optional DC preservation (see Section~\ref{ssec:perceptual}).

Figure~\ref{fig:extension} demonstrates our proposed extension for computing TIVs over higher-dimensional PTFRs (i.e., with $N=96$, $o=8$), omitting our normalization steps for visual clarity. Figure~\ref{fig:extension}a illustrates a $12$-D chroma vector, alongside $96$-D versions realized in two distinct ways: 1) \emph{zero-padding} the chroma vector with $12(o-1)$ zeros, and 2) scaling the chroma vector by $1/o$, and \emph{replicating} the result an additional $(o-1)$ times (i.e., performing a convolution with a dilated averaging kernel). Figure~\ref{fig:extension}b plots the corresponding (real/imaginary-stacked) TIV representations. Both 96-D constructions yield identical values at the frequency indices corresponding to the original chroma TIV, demonstrating that the
\begin{figure}[ht]
\centerline{\includegraphics[width=0.9\columnwidth]{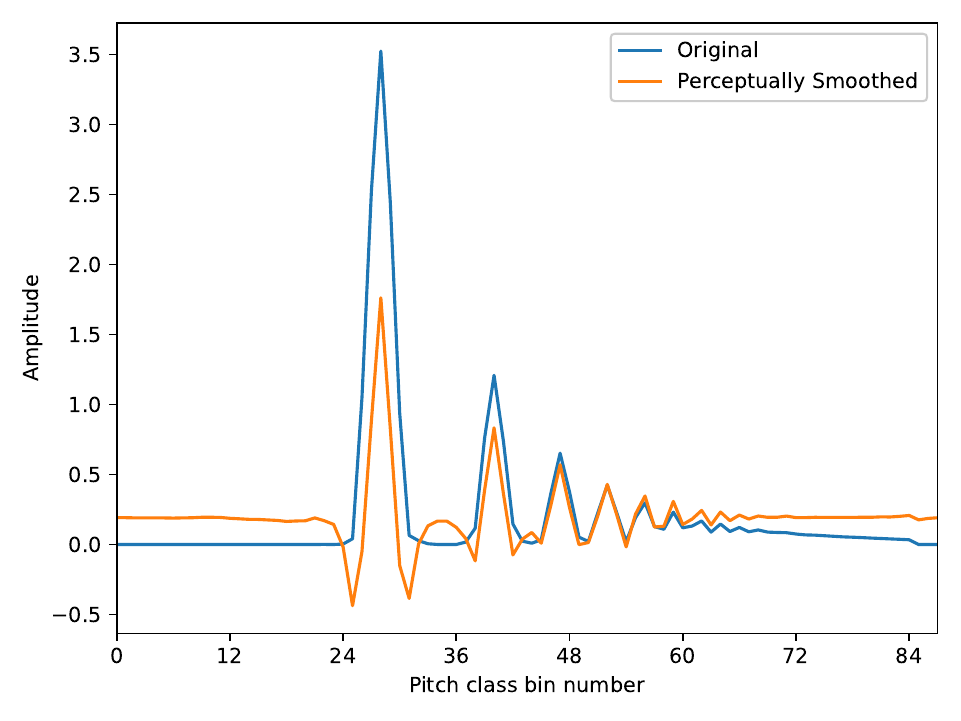}}
\caption{\label{fig:ptfr}{$88$-D PTFR and its perceptually smoothed version.}}
\vspace{-1.25 em}
\end{figure}
extension preserves the perceptual weighting defined over chroma vectors. The replicated version yields zero at all other frequencies due to the Fourier transform properties of periodic signals. This can be interpreted as octave folding acting as a blurring operation in the higher-dimensional PTFR directly. Lastly, the zero-padded version naturally convolves said spectrum with a $\operatorname{sinc}$ function.

\subsection{Perceptually smoothed PTFRs}\label{ssec:perceptual}
Lastly, we define the feature extractor $A(\cdot)$ as
\begin{equation}
A(\bm{c})=\bm{c}^\prime=\bm{F}^T\bm{Tc}=\bm{F}^T\bm{WFc}
\label{eqn:fe}
\end{equation}
where we recover $\bm{c}^\prime = \bm{c}$ for $w_*=\mathbf{1}$ and $\kappa=1$. Equation~(\ref{eqn:fe}) defines a viable $12o$-D PTFR, embedding perceptual consonance priors while admittedly no longer guaranteeing the non-negativity of $\bm{c}^\prime$, as is the case for $\bm{c}$. This representation will prove crucial in the development of our PTFR alignment algorithm in Section~\ref{sec:alignment}. We can adapt our operator to any non-$12o$-D PTFR (e.g., the $N=88$ case representing the keys of a standard piano keyboard) by appropriately zero-padding it to the nearest $12o$, applying our perceptually smoothed PTFR processing, and truncating the result to its initial dimensionality. It also scales to time-varying PTFR matrices of dimension $N \times M$. Figure~\ref{fig:ptfr} summarizes the culmination of all of our TIV extensions, illustrating an $88$-D transcription-like vector alongside its perceptually smoothed version computed using the inversion, normalization, and feature extraction techniques proposed in this section. The resulting processed version looks like its original PTFR, albeit filtered by weights in the TIV space. Since formulating our method, we observe a structural similarity between our operator and data-driven Fourier Neural Operators (FNO) \cite{li2021fourier}. The distinction is that our approach defines a fixed, perceptually grounded operator with diagonal, non-negative weights from perceptual consonance data, whereas FNOs learn broader transformations from data with respect to some objective. 

\section{Alignment of PTFRs}
\label{sec:alignment}
Given the extensions proposed in Section~\ref{sec:extensions}, we are interested in deriving a matrix $\bm{P} \in \mathcal{P}_N$ capable of aligning one non-negative PTFR $\bm{x}$ to another non-negative PTFR $\bm{y}$, where $\mathcal{P}_N$ is the set of $N \times N$ permutation  matrices. The permutation constraint maintains $\norm{\bm{Pc}}_p$ = $\norm{\bm{c}}_p \forall~\bm{c}, ~0\le p\le \infty$, where we are interested in the $p = 0$ case to ensure that pitch classes are not created or removed by its application. We frame our task as the optimization
\begin{equation}
\begin{array}{cccc}
\displaystyle
\min_{\bm{P} \in \mathcal{P}_N} & J(\bm{P})
\end{array}
\end{equation}
\begin{equation}
J(\bm{P})=\norm{\bm{T}\bm{Px}-\bm{Ty}}^2_2
\end{equation}
Note that $\bm{P}$ is applied to $\bm{x}$ along the musical pitch axis (i.e., the ``time'' domain), while the objective function is defined in the TIV space (i.e., the ``spectral'' domain). Optimizations of this form are common in audio signal processing \cite{WirtingerDAFx}, with iterative solutions often employing gradient updates derived via Wirtinger calculus. The permutation matrix constraint in our case naturally complicates carrying out such an optimization, while the use of a real spectral basis $\bm{F}$ over real-valued inputs avoids the need for complex analysis techniques used in the aforementioned works.

Expanding the norm of $J(\bm{P})$, we arrive at
\begin{equation}
J(\bm{P})=\underbrace{(\bm{Px})^{T}\bm{T}^{T}\bm{T}(\bm{Px})}_{\text{Quadratic\ Term}}-\underbrace{2(\bm{Px})^{T}\bm{T}^{T}\bm{Ty}}_{\text{Cross\ Term}}+\underbrace{\bm{y}^{T}\bm{T}^{T}\bm{Ty}}_{\text{Constant\ Term}}
\label{eqn:expansion}
\end{equation}
With $\bm{W}=\bm{I}$, the quadratic term reverts to a constant $\norm{\bm{x}}^2_2$. The resulting optimization would amount to solving a Linear Assignment Problem (LAP) \cite{kuhn1955hungarian}, whose solution we will expand on in Section~\ref{ssec:argsort}. However, since $\bm{W}$ is a non-uniform diagonal weight matrix, $\bm{T}^{T}\bm{T}$ is a full matrix that will not necessarily commute with $\bm{P}$, resulting in an NP-hard Quadratic Assignment Problem (QAP) \cite{koopmans1957assignment}. This amounts to saying that the act of perceptual filtering is dependent on the permutation which we apply onto a PTFR, since it directly dictates the tonal intervals that are present.

Since $\bm{F}$ is orthogonal, with $\kappa=0$ we can express $J(\bm{P})$ as
\begin{equation}
J(\bm{P})=\norm{\bm{T}\bm{Px}-\bm{Ty}}^2_2=\norm{A(\bm{Px})-A(\bm{y})}^2_2
\label{eqn:lap}
\end{equation}
This perspective allows us to consider an intuitive spectral matching solution \cite{levitan1987inverse}, where we approximate $J(\bm{P})$ with
\begin{equation}
J_\text{LAP}(\bm{P})=\norm{\bm{P}\bm{x}^\prime-\bm{y}^\prime}^2_2=-\underbrace{2(\bm{Px^\prime})^{T}\bm{y}}_{\text{Cross\ Term}}+\underbrace{\norm{\bm{x}^\prime}^2_2+\norm{\bm{y}^\prime}^2_2}_{\text{Constant\ Terms}}
\end{equation}
whose minimization involves solving an LAP defined over $\bm{x}^\prime$ and $\bm{y}^\prime$. We incrementally explore different methods for partially or fully carrying out this minimization.

\subsection{Circular shifts}\label{ssec:argmax}
We temporarily restrict our attention to the subclass of permutation matrices corresponding to circular shifts, since these transformations admit a direct realization in the audio domain via standard pitch-shifting techniques. In our $\operatorname{argmax}$ approach, we consider a circular shift of $\bm{x}^\prime$ such that $\operatorname*{argmax}(\bm{P}\bm{x}^\prime) = \operatorname*{argmax}(\bm{y}^\prime)$. This amounts to defining the musical root of $\bm{c}^\prime$ as the index of its maximum component (which may be violated in practice in both audio and symbolic domains, but serves as a reasonable prior in the absence of additional information). Since it does not consider the full structure of $\bm{c}^\prime$ (perceptual filtering aside), this approach often yields results similar to those obtained by applying the method directly to chroma vectors. Basic alignments of this form will be leveraged further in Section~\ref{sec:interpolation}. The optimal alignment restricted to circular shifts can be obtained via the cross-correlation method \cite{oppenheim1999signals}, a standard technique for estimating relative shifts between signals. This identifies the circular shift that maximizes the similarity between $\bm{x}^\prime$ and $\bm{y}^\prime$,  rather than relying solely on their dominant pitch classes. Notably, we apply this method not directly to chroma vectors, but rather to their perceptually smoothed versions.

\subsection{Permutations}\label{ssec:argsort}
Broadening our scope to the entire family of permutation matrices, equation (\ref{eqn:lap}) admits a global minimizer as the permutation achieving $\operatorname*{argsort}(\bm{P}\bm{x}^\prime) = \operatorname*{argsort}(\bm{y}^\prime)$. The $\operatorname{argsort}$ method follows from the \emph{rearrangement inequality} \cite{hardy1952inequalities} stating that the sum of pairwise products between two vectors is maximized when their components are sorted in the same order. As this term appears (with negative sign) as the cross term in the LAP objective, the optimal permutation aligns the order statistics of the two vectors.

Lastly, we extend our alignment approach to time-varying, non-negative PTFRs $\bm{X}$ and $\bm{Y} \in \mathbb{R}^{N \times M}$, leveraging their perceptually smoothed versions $\bm{X}^\prime$ and $\bm{Y}^\prime$, respectively. Accordingly, the true QAP and approximate LAP objectives are given by
\begin{equation}
J(\bm{P})=\norm{\bm{T}\bm{PX}-\bm{TY}}^2_F=\norm{A(\bm{PX})-A(\bm{Y})}^2_F
\end{equation}
\begin{equation}
J_\text{LAP}(\bm{P})=\norm{\bm{P}\bm{X}^\prime-\bm{Y}^\prime}^2_F =-\underbrace{2\langle\bm{PX^\prime},\bm{Y}^\prime\rangle}_{\text{Cross\ Term}}+\underbrace{\norm{\bm{X}^\prime}^2_F+\norm{\bm{Y}^\prime}^2_F}_{\text{Constant\ Terms}}
\end{equation}
respectively. The latter is identified as a LAP with cost matrix
\begin{equation}
\bm{C}_{ij}=-2\langle\bm{X}_j^\prime, \bm{Y}_i^\prime\rangle
\end{equation}
which can be solved optimally using the Hungarian algorithm \cite{kuhn1955hungarian}. The Hungarian method could also be used to minimize equation (\ref{eqn:lap}) defined over vectors, but it will converge to the same solution given by the rearrangement inequality. This can be understood through the Monge structure \cite{burkard2009assignment} of the induced cost matrix, guaranteeing the order-preserving quality in this case. In contrast to the vector case, the matrix case captures time-varying interactions between $\bm{X}^\prime$ and $\bm{Y}^\prime$ as the assignment is determined by similarities between entire feature vectors rather than a single scalar summary.
\begin{figure}[b]
\centering
  \begin{minipage}{1.0\columnwidth}
	\centering
  \centerline{\includegraphics[width=1.0\columnwidth]{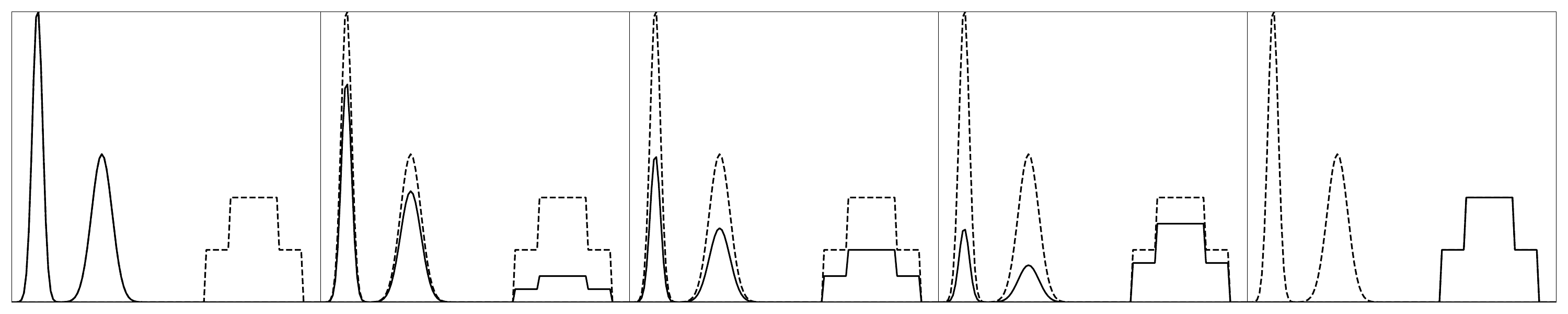}}
	\vspace{0.0em}(a)
  \end{minipage}
  \begin{minipage}{1.0\columnwidth}
	\centering
  \centerline{\includegraphics[width=1.0\columnwidth]{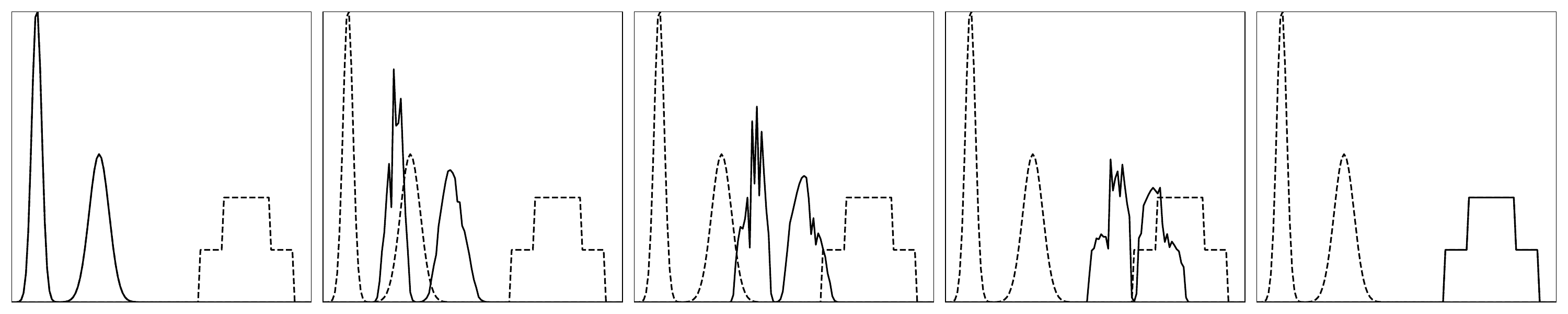}}
	\vspace{0.0em}(b)
  \end{minipage}
  \begin{minipage}{1.0\columnwidth}
	\centering
  \centerline{\includegraphics[width=1.0\columnwidth]{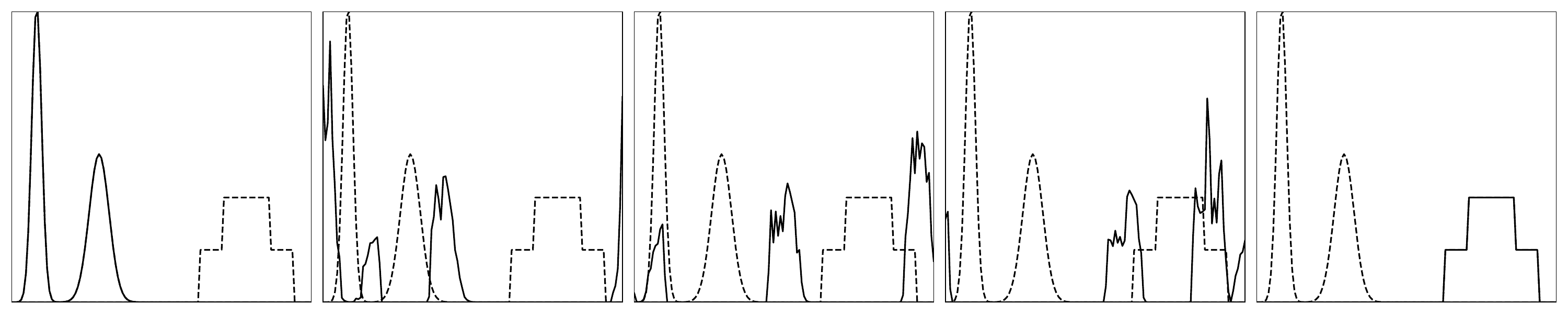}}
	\vspace{0.0em}(c)
  \end{minipage}
  \caption{(a) Naive interpolation, (b) linear and (c) circular OT.}
  \vspace{-1.5em}
  \label{fig:interp}
\end{figure}
\subsection{Iterative refinement to solve the QAP}\label{ssec:iter}
Until now, we have avoided solving the QAP, diverting our attention towards solving the related but inexact LAP. Meanwhile, we find that solving the LAP accounts for a good majority of the potential error improvement in our optimization. We suggest a 2-opt algorithm \cite{croes1958method} performing a greedy swap search to refine our initial estimate. Our iterative refinement approach works as follows:

\begin{enumerate}
    \item Find the permutation matrix $\bm{P}$ minimizing $J_\text{LAP}(\cdot)$.
    \item For every pair of indices $(i,j)$:
        \begin{enumerate}
            \item Tentatively swap the rows $\bm{P}_i$ and $\bm{P}_j$ to yield $\bm{P}^\text{new}$.
            \item Perform the update $\bm{P} \leftarrow \bm{P}^\text{new}$ if the loss improves.
        \end{enumerate}
    \item Repeat until no single swap reduces $J(\cdot)$.
\end{enumerate}
Since we are primarily interested in optimizing $\bm{P}$ for small $N$ (i.e., the $o=1$ case with $N=12$), this proves to be a pragmatic and efficient way to improve upon our initial estimate. 

%We note a natural alternative solution to the QAP using the Frank-Wolfe algorithm \cite{frank1956algorithm}, which derives a gradient update by linearizing $J(\bm{P})$ and leveraging the solution of the resulting LAP at each iteration. However, we empirically found that in this application, its necessary continuous relaxation of $\bm{P}$ leads to sub-par results when ``snapped back'' to the discrete space. In contrast, the proposed two-stage solution avoids the ``fuzziness'' of the continuous relaxation and remains in the discrete space by design.

\begin{algorithm}[t]
  \caption{Computing The Linear OT Plan $\pi^*$}\label{alg:ot}
\begin{algorithmic}
  \State{$\pi^*_{i,j} \leftarrow 0$}
  \State{$\rho_{x}, \rho_{y} \leftarrow \bar{\bm{x}}[0], \bar{\bm{y}}[0]$}
  \Comment{$\rho$ is the mass left in a bin}
  \\
  \Loop
    \If{$\rho_x < \rho_y$}
      %\State{\emph{Assign as much mass as possible}}
      \State{$\pi^*_{ij} \leftarrow \rho_X$}
      \Comment{Assign as much mass as possible}
      \\
      %\State{\emph{Refill the emptied bin}}
      \State{$i \leftarrow i+1$}
      \Comment{Refill the emptied bin}
      \If{$i \geq N$} break\EndIf
      %\State{\emph{Decrease the capacity of the other}}
      \State{$\rho_y \leftarrow \rho_y - \rho_x$}
      \Comment{Decrease the capacity of the other}
      \\
      \State{$\rho_x \leftarrow \bar{\bm{x}}[i]$}
      \\
    \Else
      \State{Symmetric to the case above}
    \EndIf
  \EndLoop
  \\\\
  \Return{$\pi^*$}
\end{algorithmic}
\end{algorithm}
\section{interpolation of PTFRs}

\label{sec:interpolation}
We aim to musically interpolate between two $L_1$-normalized PTFRs $\bar{\bm{x}}$ and $\bar{\bm{y}}$. In contrast to the alignment setting, we swap the permutation (i.e., $L_0$-preserving) constraint with a relaxed $L_1$-preserving one, maintaining the interpolated result as a valid probability vectors. A naive solution takes convex combinations of $\bar{\bm{x}}$ and $\bar{\bm{y}}$, which simply fades one set of pitches out while another fades in \cite{OTdafx}. To overcome this, we build on the approach of \cite{OTdafx}, which used OT \cite{villani2009optimal} to achieve an audio effect by interpolating reassigned spectrogram frames over a crossfade interval. While that work considered a linear frequency axis in an audio processing context, we extend it to arbitrary PTFRs aligned to a log-frequency pitch axis, and further propose a specialized circular formulation for chroma vectors. Figure~\ref{fig:interp} contrasts these methods, adapting the notional illustration from \cite{OTdafx}. Naive interpolation corresponds to a linear fade effect, while the displacement method in \cite{OTdafx} yields a more intuitive ``earth-moving'' effect. Our circular formulation utilizes both directions around the musical pitch class circle.

\begin{algorithm}[t]
  \caption{Circular OT Interpolation For ($\pi^*$, $t$)}\label{alg:cot}
\begin{algorithmic}
  \State{Given $\pi^*, t$}
  \State{$\bm{z} \leftarrow \mathbf{0}$}
  \Comment{$\bm{z}$ is the interpolated result}
  \\
  \For{$(i,j) \in \mathrm{supp}(\bm{\pi}^*)$}
    \State{$\Delta_\rightarrow,~\Delta_\leftarrow \leftarrow (j-i) \mod N,~(i-j) \mod N$}
    \If{$\Delta_\rightarrow < \Delta_\leftarrow$}
        \State{$\omega \leftarrow (i + t\Delta_\rightarrow) \mod N$} \Comment{move clockwise}
    \Else
        \State{$\omega \leftarrow (i - t\Delta_\leftarrow) \mod N$} \Comment{move counter-clockwise}
    \EndIf
    \State{$\omega_1, \omega_2 \leftarrow$ circular integer neighbors of $\omega$}
    \\
    \State{$\bm{z}[\omega_1],~\bm{z}[\omega_2] \leftarrow$ fractional split of $\pi^*_{ij}$}
    \EndFor
  \\\\
  \Return{$\bm{z}$}
\end{algorithmic}
\end{algorithm}

\begin{figure*}[htb]
\centering
  \centerline{\includegraphics[width=0.85\textwidth]{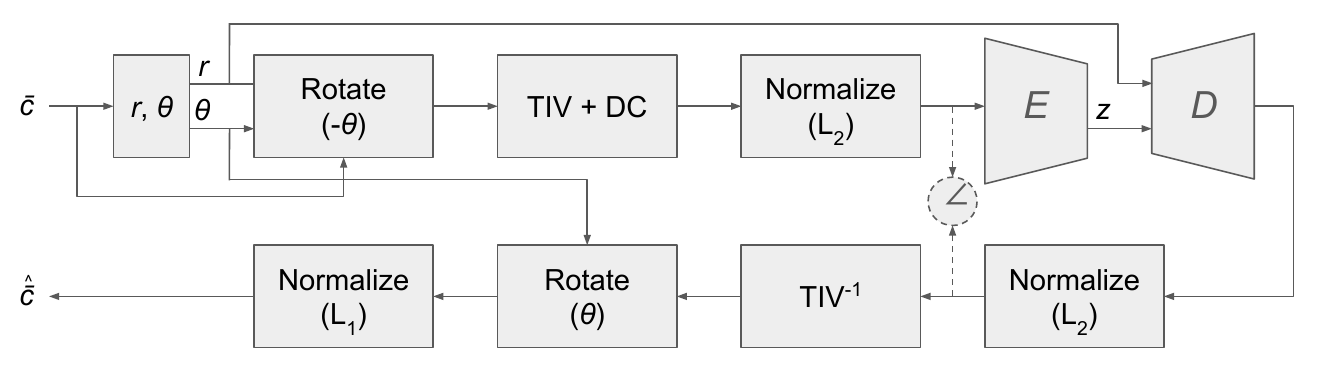}}
	\vspace{0.0em}
\caption{Block diagram of our geometric scale representation model architecture.}
\vspace{-1.0em}
\label{fig:arch}
\end{figure*}

\subsection{Linear OT}\label{ssec:lot}
The OT problem seeks the most efficient way to transform one probability distribution into another. In this setting, distributions are interpreted as mass distributions, and an OT plan $\pi^*$ describes the minimum work required to move that mass from a source $\bar{\bm{x}}$ to a target $\bar{\bm{y}}$. Viewing the pitch axis as discrete points in 1D, the OT plan is obtained via monotone rearrangement using the greedy algorithm in Algorithm \ref{alg:ot} (directly adapted from \cite{OTdafx}). We refer to this as ``linear OT'' in contrast to the circular OT variant we introduce in Section~\ref{ssec:cot}. The PTFR can be interpolated with parameter $t\in[0,1]$ by placing each mass $\pi^*_{ij}$ at
\begin{equation}
  \label{eqn:interpolated_freq}
  (1 - t)\omega_i^x + t\omega_j^y
\end{equation}
\subsection{Circle of Fourths/Fifths (CoF)}\label{ssec:cof}
The CoF \cite{levy1985theory} provides a representation in which harmonic distances between pitch classes are better captured by the number of fourth/fifth relationships separating them, rather than their chromatic distance. Mapping a chroma vector from the chromatic space to the CoF space is defined by yet another permutation matrix $\bm{H} \in \mathbb{R}^{12 \times 12}$ with inverse $\bm{H}^T$. Applying $\bm{H}$ is summarized by \texttt{c\_cof=c[0,7,2,9,4,11,6,1,8,3,10,5]}. The circular OT method in Section~\ref{ssec:cot} is invariant to whether clockwise/counterclockwise directions correspond to fourths/fifths.

\subsection{Circular OT along the CoF}\label{ssec:cot}
When considering OT-like interpolation over chroma vectors, we incorporate two  modifications. Rather than defining transport along the chromatic axis, the OT plan is constructed within the CoF domain. Moreover, we use a circular distance cost, connecting the endpoints (i.e., C and B) to enable transport over a periodic domain. As a result, the circular OT plan (readily computed using the Python OT library \cite{flamary2021pot}) can move mass in either direction around the pitch-class circle. Given the circular OT plan, interpolation is carried out using the greedy algorithm in Algorithm~\ref{alg:cot} assessing the shortest-path transfer of mass for each non-zero element $\pi^*_{ij}$. When interpolating between two chroma vectors, it is instructive to distinguish between interpolation respecting musical roots and one that captures their underlying scale color independent of root. In the latter case, we may first align one or both chroma vectors to a common root using the $\operatorname{argmax}$ method described in Section~\ref{ssec:alignment} before applying the circular OT interpolation.

\section{Geometric scale representation}\label{sec:scale}

The alignment procedure in Section~\ref{sec:alignment} enforces a discrete transformation of one chroma vector onto another under a strict $L_0$-norm preservation constraint. The interpolation procedure in Section~\ref{sec:interpolation} induces a continuous transformation under a relaxed $L_1$-norm preservation constraint. Accordingly, we introduce a scale representation learned over chroma vectors enabling interpolation along distinct dimensions of musical structure (scale root, color, and density). We are particularly interested in interpolation along the color axis while preserving the underlying scale density. Beyond providing a chroma compression effect, the approach bridges the gap between alignment and interpolation, enabling continuous musical interpolation under structured density considerations.

We formulate an autoencoder enforcing a cylindrical, $3$-D encoding, as illustrated in Figure~\ref{fig:arch}. Of its $3$ dimensions, the $r$ and $\theta$ dimensions are calculated deterministically, while the $z$ dimension is learned given access to $r$. The latter is computed as the Hoyer sparsity metric \cite{Hoyer}, scaled such that a chroma vector with a single note active maps to $1/12$, while a uniform chromatic scale maps to 1. The variable $\theta$ represents the angular distance between the determined root from C along the CoF, using the strict mathematical definition of the root as $\operatorname*{argmax}(\bm{c})$. Inspired by \cite{wu2023selfsupervised}, we leverage \emph{transposition-invariant training} \cite{arzt2018transposition}, whereby we pre-rotate chroma vectors according to $-\theta$ (such that our model only observes chroma vectors with C as their root), followed by a post-rotation by $\theta$ following the core neural network processing.

For a given pre-rotation of $\bar{\bm{c}}$, the input to the model encoder $E$ is the L2-normalized, stacked TIV representation with DC preservation (i.e., $\kappa=1$), denoted $(\bm{T\bar{c}})_{L_2}$. The encoder outputs a latent variable $z$, which is concatenated with $r$ and passed to the decoder $D$, producing a reconstruction $\widehat{(\bm{T\bar{c}})}_{L_2}$. Model optimization minimizes the cosine distance between the input $(\bm{T\bar{c}})_{L_2}$ and $\widehat{(\bm{T\bar{c}})}_{L_2}$. To recover its chroma, the reconstructed TIV is inverted and the resulting profile is rotated back to its original root by means of $\theta$. A final $L_1$ normalization yields $\hat{\bar{\bm{c}}}$. Both $E$ and $D$ are composed of $3$ linear layers with a hidden dimensionality of $2$, interleaved with LeakyReLU activations.

\begin{figure*}[hbt]
\centering
  \begin{minipage}{.45\textwidth}
	\centering
  \centerline{\includegraphics[width=0.95\columnwidth]{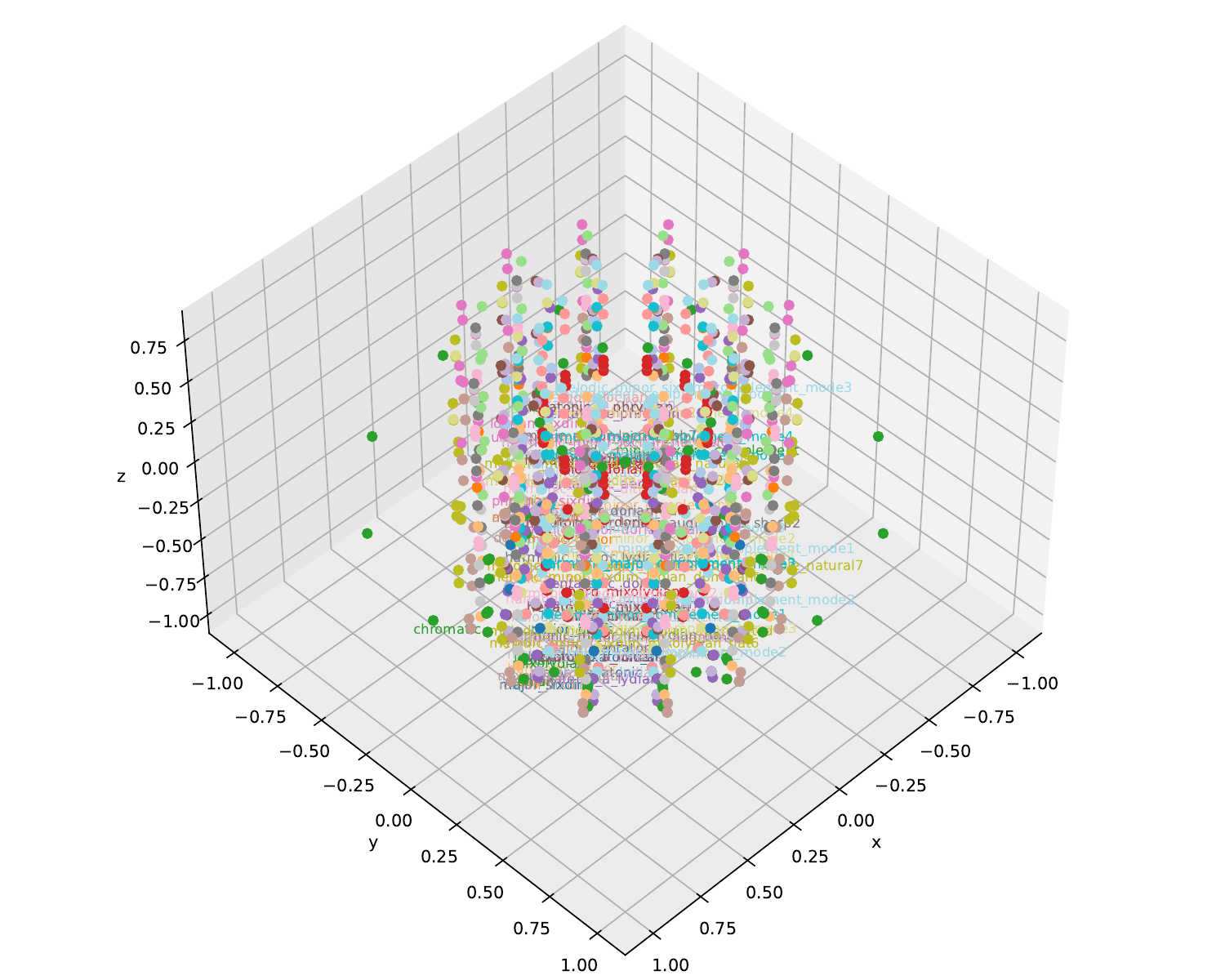}}
	\vspace{0.0em}(a)
  \end{minipage}
  \begin{minipage}{.45\textwidth}
	\centering
  \centerline{\includegraphics[width=0.95\columnwidth]{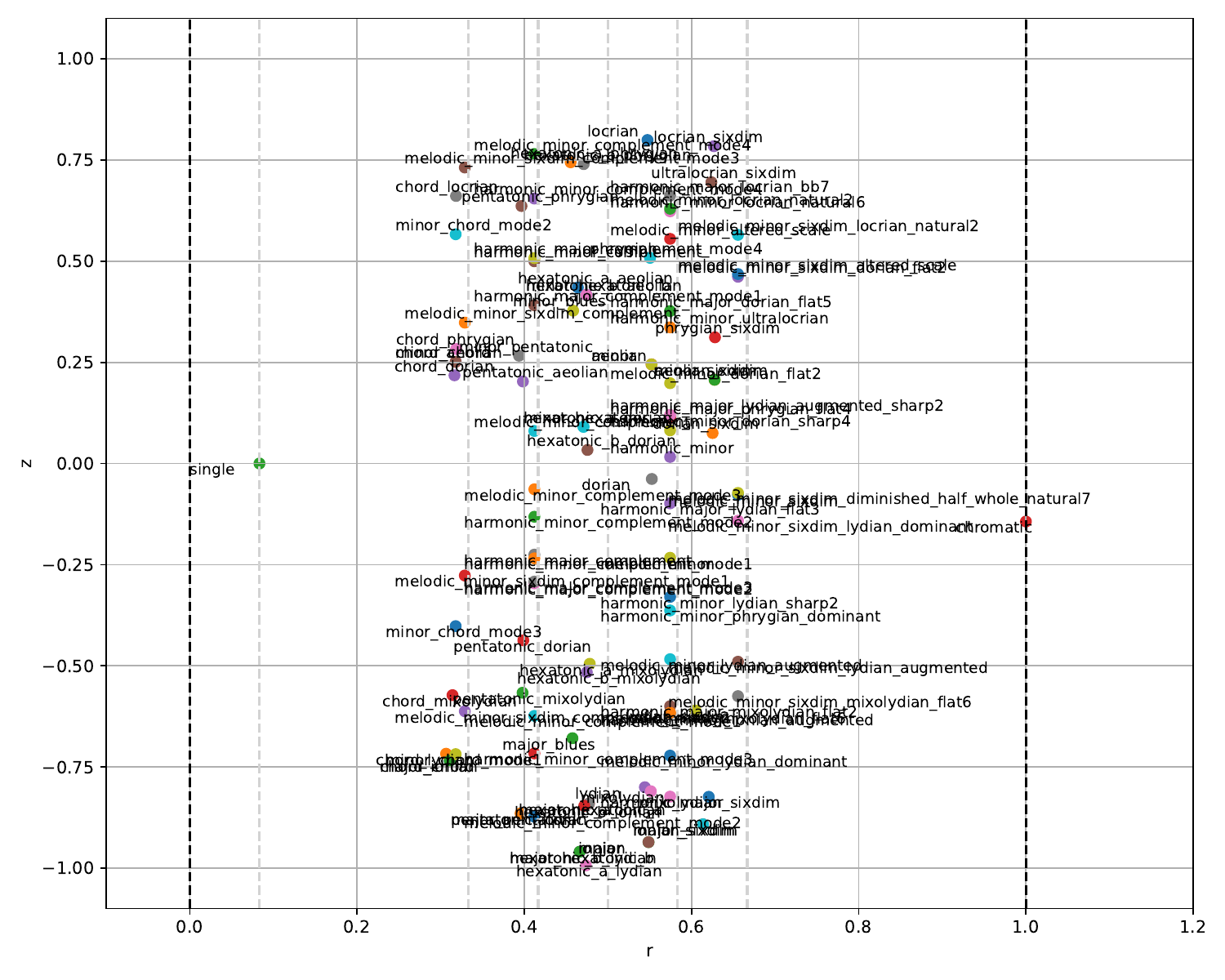}}
	\vspace{0.0em}(b)
  \end{minipage}
\caption{(a) Cylindrical scale representation and (b) rectangular planar profile for $\theta=0$.}
\vspace{-1.5em}
\label{fig:scale_rep}
\end{figure*}

\setlength{\tabcolsep}{1.5pt}
\begin{table}[ht]
\caption{Sequence alignment results, with random columns reporting averages over 1024 randomly generated chromagram pairs.}
\centering
\begin{tabular}{lcccc}
\toprule
&
\multicolumn{2}{c}{$\norm{T(\bm{Px}_{\text{PCP}})-T(\bm{y}_{\text{PCP}})}_2$}
&
\multicolumn{2}{c}{$\norm{T(\bm{PX})-T(\bm{Y})}_F$}
\\
\cmidrule(lr){2-3}
\cmidrule(lr){4-5}
\textbf{Method}
& Musical
& Random
& Musical
& Random
\\
\midrule
None
& $23.354$
& $7.308$
& $130.450$
& $26.344$
\\
$\operatorname{argmax}$
& $13.206$
& $6.534$
& $104.226$
& $24.174$
\\
Cross-correlation
& $13.206$
& $5.055$
& $104.226$
& $20.277$
\\
$\operatorname{argsort}$
& $5.361$
& $3.086$
& $100.266$
& $15.964$
\\
$\operatorname{argsort}$+2-opt
& $\mathbf{4.350}$
& $\mathbf{2.276}$
& $107.568$
& $14.509$
\\
Hungarian
& $7.622$
& $3.183$
& $90.537$
& $14.889$
\\
Hungarian+2-opt
& $7.891$
& $2.559$
& $\mathbf{89.847}$
& $\mathbf{13.012}$
\\
\bottomrule
\end{tabular}
\label{tab:synthetic}
\vspace{-1em}
\end{table}

\section{Experimental results}\label{sec:results}
Due to space constraints, we refer readers to additional figures and audio demos available at \url{https://tiv-ext.netlify.app}.

\subsection{Alignment of PTFRs}\label{ssec:alignment}
To illustrate our proposed alignment approach, we consider a simple example consisting of two synthetically generated \emph{musical} signals. The chromagram $\bm{X}$ represents a I-IV-V-I chord sequence in D major, with slight perturbations applied to its weights for realism. The chromagram $\bm{Y}$ represents the standard minor cantus firmus $1$-$5$-$4$-$\flat3$-$2$-$1$-$\flat3$-$2$-$1$ \cite{fux1965study} in C minor. We derive permutations using the methods outlined in Section~\ref{sec:alignment}. For methods operating over a single chroma vector, we compute PCPs ${\bm{x}}_{\text{PCP}}$ and ${\bm{y}}_{\text{PCP}}$ for $\bm{X}$ and $\bm{Y}$, respectively. The resulting permutation is then applied to $\bm{X}$.
In order to demonstrate our alignment approach at scale, we also create a \emph{random} dataset of 1024 randomly generated pairs of $\bm{X}$ and $\bm{Y}$ chromagrams to assess average optimizer performance.

Alignment results are summarized in Table \ref{tab:synthetic}, illustrating the incremental improvements incurred between methods. In order to avoid biasing results towards our own creations, distance metrics are reported given the original TIV in equation (\ref{eqn:tiv}). We see that the $\operatorname{argsort}$ and Hungarian methods yield the best results for the LAP approximation for the vector and matrix cases, respectively. The proposed 2-opt refinement can improve upon these solutions by solving the true QAP. Though less expressive, shift-only methods still incrementally improve upon initial alignments.

\subsection{Interpolation of PTFRs}\label{ssec:interpolation}
To demonstrate our proposed interpolation approach, we consider blending between $\bm{\bar{x}} = [1/3, 0, 0, 0, 1/3, 0, 0, 1/3, 0, 0, 0, 0]$ (C major) and $\bm{\bar{y}} = [1/3, 0, 0, 1/3, 0, 0, 0, 1/3, 0, 0, 0, 0]$ (C minor). We evaluate the resulting interpolation for $t=[0, 1/3, 1/2, 2/3, 1]$ achieved by naive linear interpolation (fading), linear OT displacement, and our proposed circular OT in the CoF domain. The baseline methods converge to the same solution in this case since the dominant transport occurs from E to D$\sharp$/E$\flat$. While these methods produce musically unstructured fades, the proposed interpolation outlines a plausible sequence of harmonic modulations, as summarized in Table \ref{tab:cot}. At $t=0.5$ (the Wasserstein barycenter \cite{agueh2011barycenters}), the interpolation yields a ``colorless'' hexatonic note set that is neither major nor minor, corresponding to the largest subset of the major scale or its modes exhibiting this property. Its invariance under reflection about the \emph{negative harmony axis} \cite{levy1985theory} highlights its intrinsic harmonic balance.
\begin{table}[t]
  \caption{Interpolation between C major and C minor triads.}
	\centering
\begin{tabular}{c|ll|ll}
\toprule
 & \multicolumn{2}{c|}{\textbf{Linear Fade / OT}} & \multicolumn{2}{c}{\textbf{Circular OT along CoF}} \\
$t$ & \textbf{Note Set} & \textbf{Chord} & \textbf{Note Set} & \textbf{Chord} \\
\midrule
$0$ & [C, E, G] & C & [C, E, G] & C \\
$1/3$ & [C, D$\sharp$, E, G] & Cadd$\sharp$9  & [C, F, G, A] & Fadd9 \\
$1/2$ & [C, D$\sharp$, E, G] & Cadd$\sharp$9 & [C, D, F, G, A, B$\flat$] & Gm9 / C \\
$2/3$ & [C, D$\sharp$, E, G] & Cadd$\sharp$9 & [C, D, G, B$\flat$] & Gm / C \\
$1$   & [C, E$\flat$, G] & Cm & [C, E$\flat$, G] & Cm \\
\bottomrule
\end{tabular}
	\label{tab:cot}
\vspace{-1.5em}
\end{table}
\subsection{Geometric scale representation}\label{ssec:scale}
We train our representation on a set of 106 internally constructed scale profiles. The profiles span tetrachord, pentatonic, hexatonic, heptatonic, and octatonic scales, as well as a chromatic and “single-note” scale with one active pitch class. Weights are derived from the Krumhansl–Schmuckler and Temperley profiles \cite{temperley1999whats} setting non-scale tones weights to zero. Each training step uses the full training set as a batch. The model is trained for 20K steps using the Adam optimizer with a learning rate of $10^{-3}$.

Figure~\ref{fig:scale_rep} shows our learned encodings over the scale set. We center $z$ using its value corresponding to the ``single-note'' scale and rescale its distribution to lie within $[-1, 1]$. We observe Locrian-like and Lydian-like profiles at the extremes of $z$, indicating that it captures a spectrum of bright and dark tonalities. Scales are organized by density along the orthogonal $r$ axis. We demonstrate smooth aural transitions of decoded scale profiles along the scale-color axis by varying $z$ while holding $r$ fixed on our demo website.

\section{Conclusions}\label{sec:conclusions}
We presented a framework for modeling pitch structure in audio and symbolic music through perceptually grounded transformations of PTFRs. By extending the TIV into an invertible and flexible operator, we enabled a new family of PTFRs and leveraged its properties for aligning musical structures. We introduced a circular OT-based interpolation that yields musically meaningful transitions between pitch distributions, consistent with principles of musical voice leading. Finally, we introduced a learned geometric scale representation that combines these perspectives to create an interpretable structure in terms of scale color and density. These contributions demonstrate that perceptually informed signal processing techniques can yield interpretable models of musically pitched material, offering a foundation for structure-aware audio analysis and transformation. Future work will explore alternative alignment objectives (e.g., cosine or $L_1$ distances) and investigate the integration of our feature extractor within the circular OT framework, which is currently constrained by OT’s non-negativity requirement. We also aim to further develop the geometric scale representation by introducing a uniform, bounded bottleneck for $z$ (e.g., via a variational autoencoder with a truncated Gaussian prior \cite{zhao2019tgmvae} and/or finite scalar quantization \cite{mentzer2023fsq}). Finally, we plan to extend these ideas toward enabling deeper and more structured control of PTFR-conditioned generative music systems.

%\newpage
\bibliographystyle{IEEEtranDAFx}
\bibliography{DAFx26_tmpl} % requires file DAFx26_tmpl.bib

\end{document}